# Does everyone have a price? Understanding people's attitude towards online and offline price discrimination

**Joost Poort**
*Institute for Information Law (IViR), University of Amsterdam, The Netherlands, poort@uva.nl*

**Frederik J. Zuiderveen Borgesius**
*Digital Security group (DiS), Radboud University, Nijmegen, The Netherlands, f.j.zuiderveenborgesius@uva.nl*

**Abstract:** Online stores can present a different price to each customer. Such algorithmic personalised pricing can lead to advanced forms of price discrimination based on the characteristics and behaviour of individual consumers. We conducted two consumer surveys among a representative sample of the Dutch population (N=1233 and N=1202), to analyse consumer attitudes towards a list of examples of price discrimination and dynamic pricing. A vast majority finds online price discrimination unfair and unacceptable, and thinks it should be banned. However, some pricing strategies that have been used by companies for decades are almost equally unpopular. We analyse the results to better understand why people dislike many types of price discrimination.

### Article information

**Received:** 11 Aug 2018 **Reviewed:** 26 Nov 2018 **Published:** 30 Jan 2019
**Licence:** Creative Commons Attribution 3.0 Germany
**Funding:** The work of Zuiderveen Borgesius for this research was largely funded by an EU Marie Curie research grant (nr. 748514, 'Profile').
**Competing interests:** The author has declared that no competing interests exist that have influenced the text.

**URL:**
http://policyreview.info/articles/analysis/does-everyone-have-price-understanding-peoples-attitude-towards-online-and-offline

**Citation:** Poort, J. & Zuiderveen Borgesius, F. J. (2019). Does everyone have a price? Understanding people's attitude towards online and offline price discrimination. *Internet Policy Review*, *8*(1). DOI: 10.14763/2019.1.1383

## 1. INTRODUCTION

An online store can offer each website customer a different price based on his or her individual characteristics or assumed characteristics and behaviour. For instance, stores can categorise consumers according to their presumed wealth – or rather: price-sensitivity – and charge people that are less price-sensitive higher prices. Such *personalised pricing* is an online form of price discrimination. For the scope of this paper, personalised pricing is defined as: *differentiating the online price for identical products or services based on information a company has about a potential customer*. Online price discrimination is an example of a broader trend towards data-driven or algorithmic personalisation of services. Several commentators expect that online price discrimination will become more prevalent in the near future (e.g., Odlyzko, 2009; Executive Office of the President of the United States, 2015).

This paper focuses on the question of how people feel about online price discrimination and about several other forms of price discrimination and dynamic pricing based on other factors such as shifts in demand. The paper aims to gain a better understanding of the drivers of people's acceptance or rejection of price discrimination and dynamic pricing in different settings, and the economic and demographic factors underlying it. To this end, we conducted two surveys in the Netherlands, among a representative sample of the Dutch population (N=1233 and N=1202). An overwhelming majority considers online price discrimination unacceptable and unfair. However, people also dislike some pricing strategies that have been commonly applied for decades.

To the best of our knowledge, this is the first detailed survey in a European country on people's attitudes towards different forms of price discrimination. Outside Europe such surveys are rare too (with Turow, Feldman and Meltzer (2005) and Turow, King, Hoofnagle, Bleakley, and Hennessy (2009) as notable exceptions, in the US). The paper could be relevant for policymakers that consider regulating online price discrimination, and for companies that consider applying it.

This paper is structured as follows: section 2 gives an introduction to online price discrimination, and section 3 discusses the basic economics of price discrimination and summarises the legal status of online price discrimination in Europe. The setup and results of the consumer surveys are presented and discussed in section 4. Demographic patterns are analysed in section 5. Section 6 offers concluding thoughts.

## 2. ONLINE PRICE DISCRIMINATION

Several examples of online price discrimination have been documented. Almost twenty years ago, Amazon reportedly charged more to existing customers than to would-be customers (BBC News, 2000). When a regular customer deleted his computer's cookies, he saw the price of a DVD drop. Hence, it appeared that customers who previously ordered from Amazon were charged more for a product than new customers. When this caught media attention, Amazon hastily issued a press release stating that it was merely experimenting with random discounts and gave a refund to people who paid a price above the average. Amazon's CEO Jeff Bezos said: "We have never tested and we never will test prices based on customer demographics" (Amazon, 2000).

The first hard evidence of online stores adapting prices to customers is from 2012. The US office supply store Staples adapted prices to the area where customers were based, on the basis of the customers' IP addresses – unrelated to shipping costs (Valentino-DeVries, Singer-Vine & Soltani, 2012). Because of this pricing scheme, the store charged lower prices to people from high-income areas. Presumably, that effect was unintended. Other stores also adapt prices to the customer's area (Mikians, Gyarmati, Erramilli, & Laoutaris, 2013). And some stores are reported to offer discounts to customers who use a mobile device or that are logged in (Hannak, Soeller, Lazer, Mislove, & Wilson, 2014). Many people suspect that airlines adapt ticket prices to the customer's browsing activities. There is, however, no evidence for such practices (Vissers, Nikiforakis, Bielova, Joosen, 2014).

It is clear that technology for personalised pricing is available. Companies can adapt online advertising to individual internet users – a practice called behavioural advertising – and could use the same technology to adapt prices. If online price discrimination can be used to legally increase profits, companies can be expected to do so.

On the other hand, stores may be more careful about mistakes in the context of personalised pricing than in that of behavioural advertising. Targeting a football-loving customer with an ad for a hockey stick may not scare away that customer. But showing a high price to a customer with a low willingness to pay may do just that. Stores may also fear that consumers who find out about price discrimination become angry (Odlyzko, 2009).

## 3. ECONOMIC BACKGROUND AND LEGAL STATUS [1]

### 3.1. DEFINITION AND TYPES OF PRICE DISCRIMINATION

As noted, online price discrimination can be defined as differentiating the online price for identical products or services based on information a company has about a potential customer. This definition is also used in Zuiderveen Borgesius and Poort (2017). The word 'identical' is crucial here, as this sets aside price differentials that are induced by variations in the *costs* of serving various customers. Think for instance of different shipping costs or different risk profiles in insurance and credit markets. Based on demographic data or a person's track record, he or she may have a higher probability to cause a traffic accident, fall ill, become unemployed, or default on a loan. By consequence, the cost of providing insurance or credit will differ. These cost differences justify price differences that most authors would not consider price discrimination. Versioning, by means of which similar but not identical products are sold at different markups, also falls outside the scope of this paper. [2] Simple examples of online price discrimination are a reduced conference fee for academics or doctoral students (and a higher fee for participants from commercial entities), or a reduced fee for children when booking theatre or airline tickets online.

For price discrimination to work, three conditions must be satisfied: (i) The seller must be able to distinguish between customers to know which price to charge to whom; (ii) The seller must have enough market power to be able to set prices above marginal costs; (iii) Resale must be impractical, costly, or forbidden to prevent arbitrage between customers (e.g., Varian, 1989). For online sales these conditions are often met: distinguishing customers is possible with great accuracy. Various internet sellers have very high market shares in their relevant market which are likely to give them at least some market power, while market power may also derive from switching costs or lack of transparency in the market. And resale is often impossible (in case of airline tickets or hotel rooms for instance) or relatively costly. Combined with the ease of adapting prices unnoticed, one might expect a lot of price discrimination to occur online.

A classic distinction is between first, second, and third degree price discrimination (Pigou, 1932). *First degree price discrimination* refers to a situation in which each consumer is charged an individual price equal to his or her maximum willingness to pay. This requires precise information about the buyer's willingness to pay, the reservation price. First degree price discrimination enables sellers to extract all consumer surplus. In practice, such an extreme form of price discrimination will never occur, as sellers cannot learn buyers' exact reservation price. First degree price discrimination serves as a stylised benchmark to evaluate other pricing schemes. [3]

*Second degree price discrimination* refers to pricing schemes in which the price of a good or service depends on the *quantity* bought. Such schemes are also called 'non-linear pricing' and may involve a quantity discount, or a two-part tariff with a fixed fee and a variable fee. For example, in the cinema popcorn is often cheaper (per gram) if you buy a larger box. For second degree price discrimination the seller does not need information about the buyer, as buyers self-select: they choose a different price by choosing a different quantity.

Loyalty schemes are sometimes also characterised as second degree price discrimination. This is correct if a loyalty scheme only amounts to a quantity discount over time: past purchases giving a discount on future purchases. However, loyalty schemes are often used to sell additional or more profitable products and services to existing customers (cross-selling or up-selling). Therefore, by building customer profiles, loyalty schemes can also be used for personalised pricing. In such cases loyalty schemes should rather be seen as third degree price discrimination.

In *third degree price discrimination*, prices differ between groups or types of buyers. This type of price discrimination is widely used: discounts for students, children or elderly are well-known examples. A company could also charge people from different geographical areas different prices. For instance, medicines or college textbooks could be sold at lower prices in developing countries.

For third degree price discrimination it is not necessary to recognise individual buyers: Sellers only need to know the characteristics of the buyer that are used to discriminate prices. However, to distinguish types of buyers, sellers often use unique identifiers such as a student card with a student number and photo or even a formal ID-card. Uniquely identifying customers helps to satisfy two of the key conditions for price discrimination to work: distinguishing between buyers and preventing ineligible customers to obtain a discount by arbitrage.

Online price discrimination will typically work similarly: an online store identifies a customer on the basis of, for instance, a cookie, an IP-address, or user log-in information. Like the student ID, this unique identification will generally not be the purpose but a means to the end of third degree price discrimination by distinguishing between broader categories, for instance high and low spenders. However, compared to selecting students on the basis of a student ID card, an online profile can be much more detailed and can allow for much more refined price discrimination. By doing so, online third degree price discrimination can, at least in theory, be pushed towards a seller's holy grail of perfect or first-degree price discrimination, under which all consumer surplus is extracted to the benefit of the seller.

A pricing strategy which is related to price discrimination and which is often used on the internet is *dynamic pricing* or *time-based pricing*. Under dynamic pricing, a company adjusts prices based on market conditions concerning supply and demand. An airline company, for instance, will generally raise the price of tickets for a flight if it is almost fully booked. Similarly, it will charge higher prices at popular times and days, for instance for tickets to beach destinations during school holidays.

## 3.2. WELFARE EFFECTS OF PRICE DISCRIMINATION

Price discrimination can benefit both buyers and sellers, leading to an increase of both consumer and producer welfare (for an illustration of this, see Zuiderveen Borgesius and Poort, 2017: p. 353-354). It can help the seller to regain its fixed costs without leaving many potential customers unserved. The same is true for dynamic pricing.

On the other hand, for some customers price discrimination and dynamic pricing will lead to higher prices than uniform or constant prices. Hence, such pricing strategies deprive some consumer groups of welfare (consumer surplus). The more refined the pricing scheme that the seller uses, the more this can be the case.

There is a large literature on the outcomes and welfare effects of price discrimination in various different competitive settings and under various assumptions about consumer demand, information that consumers and producers have, producers' ability to commit to prices, etc. For an overview, see for instance Varian (1989) and Armstrong (2006). These welfare effects turn out to be ambiguous. When price discrimination does not lead to substantial market expansion, it often reduces total consumer surplus to the benefit of producer surplus. Price discrimination may even lead to a net welfare loss, when producers gain but consumers lose more. And sometimes even sellers can suffer a welfare loss, due to intensified competition.

Generally, for price discrimination to be welfare enhancing, it must lead to a substantial increase in total output by serving markets that were previously unserved. But even then, consumers with a high willingness to pay will most probably be worse off under price discrimination and the closer personalised pricing approaches first degree price discrimination, the more it will extract welfare away from consumers and towards producers.

## 3.3. LEGAL STATUS IN EUROPE

In Europe, there are no specific laws on online price discrimination. Scholars are beginning to explore whether existing legal principles can help to mitigate possible consumer harm resulting from online price discrimination.

Data protection law, in particular the General Data Protection Regulation (2016), or GDPR, does not contain specific rules on price discrimination. In earlier work, we analysed the relevance of the GDPR for online price discrimination (Zuiderveen Borgesius & Poort 2017). We concluded that the GDPR could help to make online price discrimination more transparent. The GDPR applies when personal data are processed. As most types of online price discrimination involve using personal data, the GDPR applies to most types. When a company uses personal data, the company must disclose the purpose of that data use. Hence, if a company uses personal data (such as log-in information, a tracking cookie, or an IP address) to recognise customers and to adapt prices, the company must disclose that it uses personal data for price discrimination. Moreover, in most cases, the GDPR probably requires companies to ask the customer's prior consent for price discrimination. Apart from the GDPR, if a company uses a cookie (or similar file) to recognise somebody, the ePrivacy Directive (2009) requires the company to inform the person about the cookie's purpose. See Zuiderveen Borgesius & Poort (2017) for a more detailed analysis of online price discrimination in relation to GDPR. See also Steppe (2017) and Zuiderveen Borgesius (2015).

European consumer law does not explicitly prohibit or regulate personalised pricing. But consumer law could be interpreted as also requiring companies to disclose that they use online price discrimination (see in particular article 5(1)(c) of the Consumer Rights Directive 2011; Neppelenbroek, 2016). Non-discrimination law may prohibit price discrimination if it harms people with certain protected characteristics, such as skin colour or gender (Art. 21, Charter of Fundamental Rights of the European Union). And, under certain circumstances, EU law prohibits price discrimination if it leads to discrimination of people from other European member states (article 4 Geo-Blocking Regulation 2018/302; Schulte-Nölke et al., 2013).

In conclusion, there are no specific Europe-wide laws regarding online price discrimination, and existing rules do not, in general, prohibit the practice. But data protection law can be interpreted as requiring transparency regarding most types of personalised pricing, and even as requiring the consumer's prior consent.

# 4. SURVEY RESULTS

## 4.1. SURVEY SETUP

To assess consumers' attitudes towards various forms of price discrimination and dynamic pricing, we conducted two surveys amongst a representative sample of the Dutch population aged 18 and older. Both surveys were held within the LISS panel. 4

The field work for the first survey was conducted in April 2016. This survey had a response of 1,233 completes for the questions relevant to this paper (81.0%). The second survey was conducted in November 2016 and had a response of 1,202 relevant completes (82.2%). Both surveys not only contained questions concerning price discrimination, but also covered other topics, mostly related to media consumption.

In the first survey, the relevant questions focused on consumers' general experience with online price discrimination and their general attitudes towards it. The survey questions did not use the term 'discrimination' or 'price discrimination' which may be normatively loaded. The survey described personalised pricing as follows: '*Web shops can adjust prices on the basis of data about an Internet user, such as the country where the user is based, or the time the user visits the web shop. This makes it possible that two Internet users, who visit the same web shop at the same time, see different prices for the same product*.' 5 After this introduction, respondents were asked about their experiences and attitudes concerning this.

In the second survey, six months later, we presented respondents with fifteen current or fictitious examples of price discrimination or dynamic pricing by web stores or offline stores – again without using such normatively loaded terms – and asked them to indicate how acceptable this practice was to them.

### 4.2. GENERAL EXPERIENCES AND ATTITUDES

After the general introduction of online price discrimination, respondents were asked in the first survey how often they had experienced it. 56.9% indicate they never experienced online price discrimination while 4.3% claim to have experienced it often or very often (see Figure 1). This implies that the subsequent questions in this first survey may have been fairly abstract and hard to assess for a number of respondents.

Next, respondents were asked to indicate on a 7-point Likert scale whether they thought such practices should be prohibited. As can be seen from Figure 2, a large majority would favour a prohibition. More than 72% of respondents choose 5, 6 or 7.

Subsequently, respondents were asked to indicate on a 7-point Likert scale whether in their eyes online price discrimination is acceptable and fair. More than 80% consider it to some extent unacceptable and unfair (Figure 3).

When the question was framed differently in terms of a *discount*, the acceptation increased somewhat, but around 65% still finds online price discrimination to some extent unacceptable and only 16-17% finds it acceptable. Respondents' acceptance hardly depends on whether price discrimination favours themselves or others (Figure 4).

Finally, to find out more about why people do or do not approve of online price discrimination, respondents were asked to indicate on a 7-point Likert scale to what extent they agreed or disagreed with three propositions about online price discrimination. Almost 80% agrees with the proposition that web stores should be obliged to inform customers if they price discriminate (Figure 5).

When asked whether they worry about paying more than others or about not noticing price adjustments, opinions are distributed more evenly (Figure 6). About 20% is neutral towards both propositions, while between 56% and 65% is, to a certain extent, worried. Generally, respondents are somewhat more worried about not noticing online price discrimination than about paying more than others.

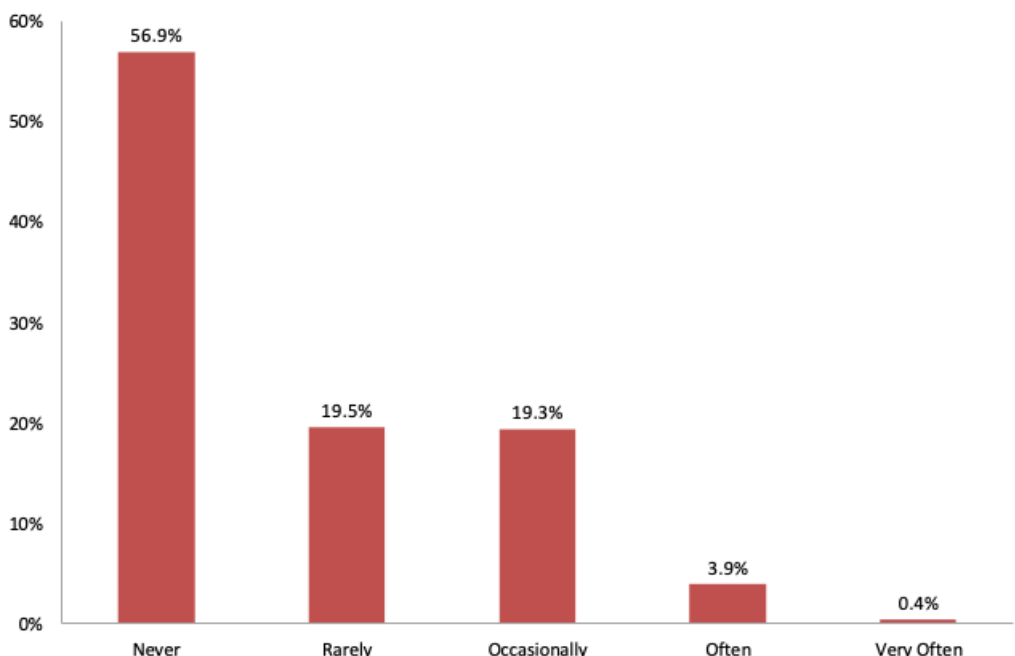

Figure 1: 'How often have you experienced [online price discrimination]?' (N=1233)

Does everyone have a price? Understanding people's attitude towards online and offline price discrimination

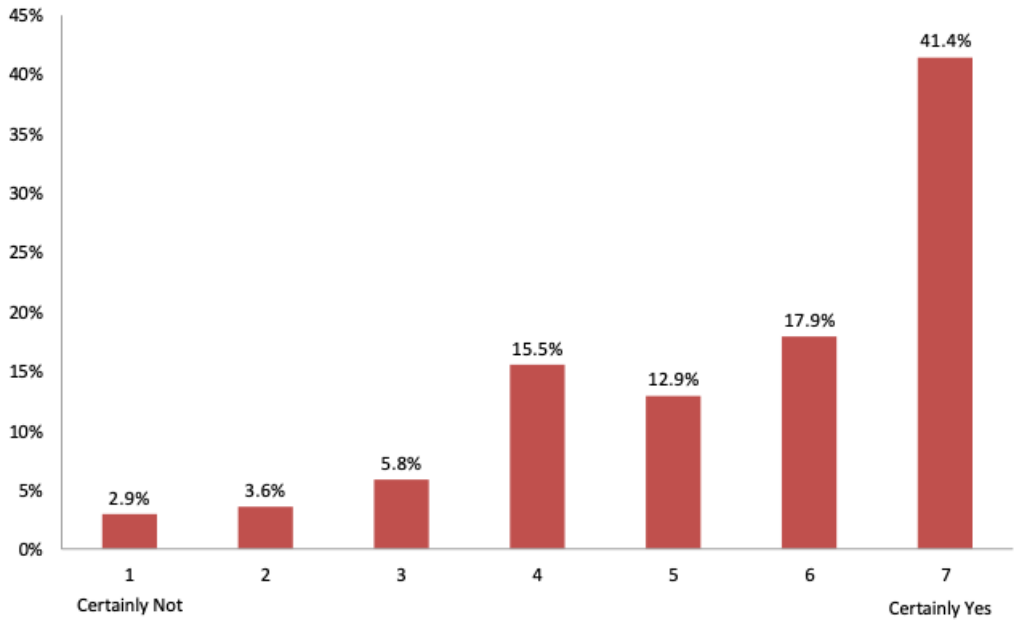

Figure 2: 'According to you, should such practices be prohibited?' (N=1233)

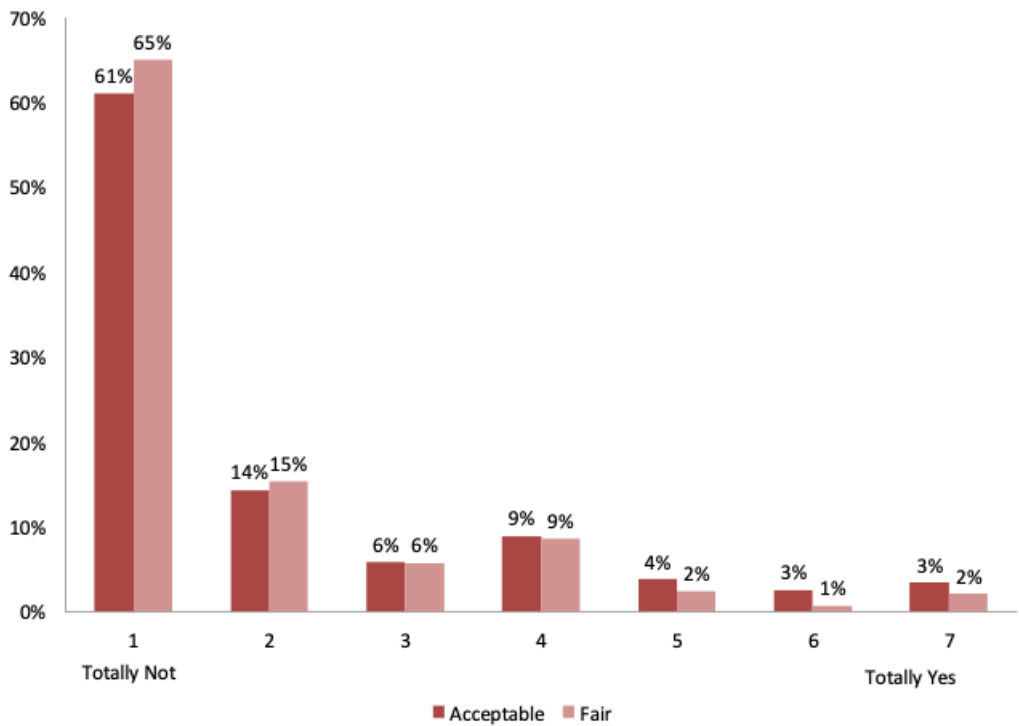

Figure 3: 'Do you find it acceptable/fair that an online store charges different prices to different people for the same product?' (N=1233)

Does everyone have a price? Understanding people's attitude towards online and offline price discrimination

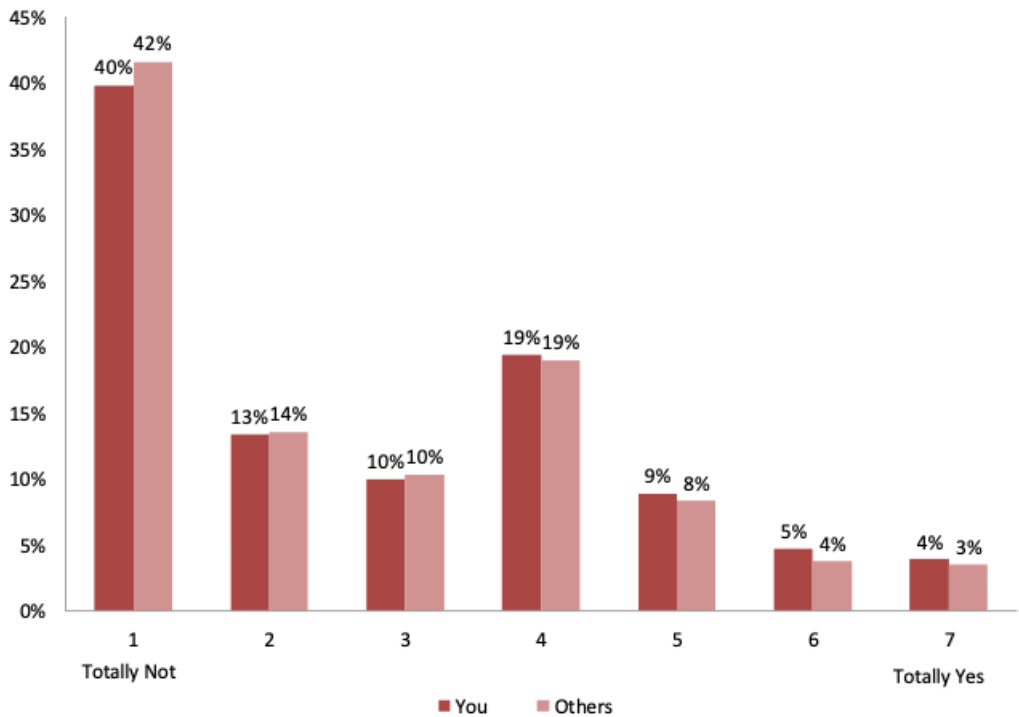

Figure 4: 'Would you find it acceptable if a web store gives a discount to you/others based on your/their online behaviour (such as the websites you/they have visited before)?' (N=1233)

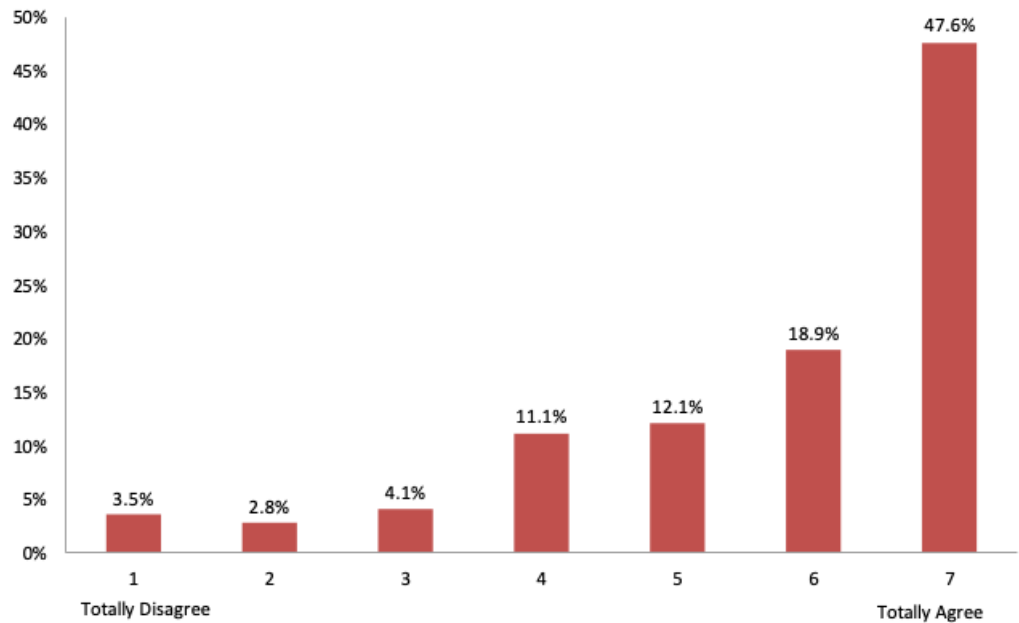

Figure 5: 'Web store should be obliged to inform customers if they charge different prices to different visitors, for the same product' (N=1233)?

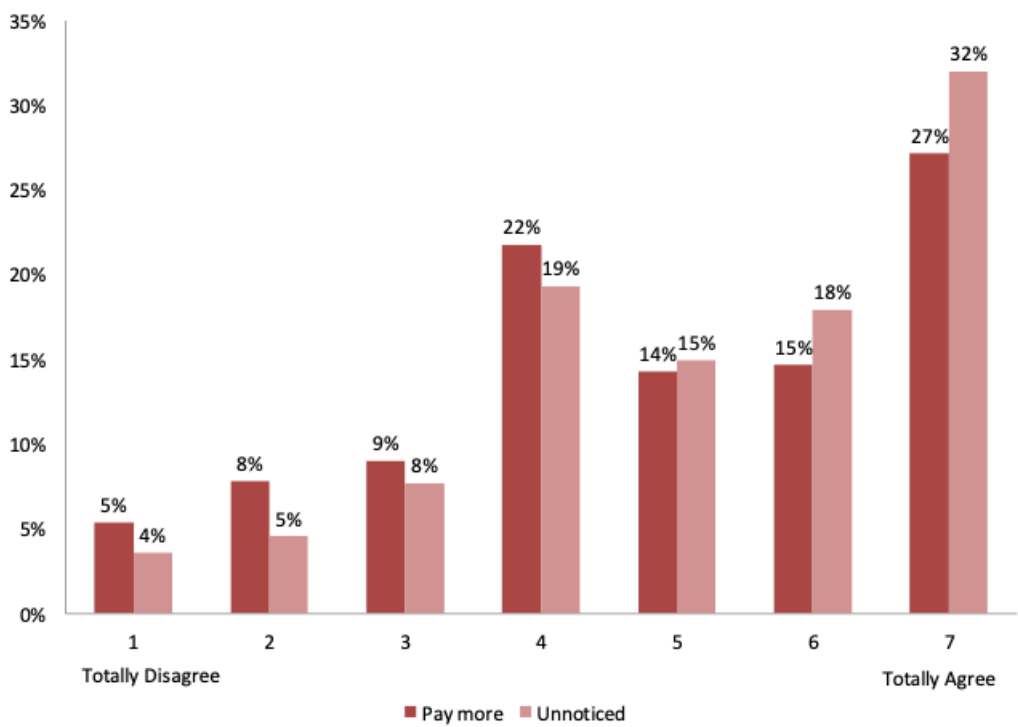

Figure 6: 'I worry that I may pay more for a product than others / price adjustments can be made without me noticing it' (N=1233)?

## 4.3. ATTITUDES TOWARDS SPECIFIC EXAMPLES

The second survey was much less generic and asked respondents about specific examples: Which sorts of online or offline price discrimination or dynamic pricing do people find acceptable? Respondents were presented with fifteen examples and, as in the first survey, were asked to indicate on a 7-point Likert scale to what extent they considered these acceptable.

Figure 7 presents the distribution of the answers for the cases respondents find the most acceptable – a discount for supermarket customers holding a loyalty card – and the most unacceptable – a higher price for hotel rooms when using an Apple computer. Loyalty cards are acceptable to the majority and 44% find them (very) acceptable (6 or 7), against 12% (very) unacceptable (1 or 2). In contrast, only 2% finds it (very) acceptable for a hotel booking site to price discriminate based on someone's type of computer, while 77% considers this (very) unacceptable.

Figure 8 summarises the responses to all fifteen questions, by presenting the 'net acceptability'. We define 'net acceptability' as the difference between the percentage indicating 6 or 7 ((very) acceptable) minus the percentage indicating 1 or 2 ((very) unacceptable). Hence, if 40% find an example (very) acceptable and 30% (very) unacceptable, the net acceptability is 40% – 30% = 10%. In Figure 8, we plotted this net acceptability from the highest to the lowest number. It ranges from +32% to –75%. The number inside the bars gives the simple average of the scores indicated by respondents. These averages range from 5.0 to 1.9 (4 being neutral) and correlate very strongly with the net acceptability.

In general, Figure 8 illustrates that people dislike many forms of price discrimination and dynamic pricing. For nine out of fifteen examples, the average score is below 4 and the net acceptability clearly below 0%. Only three examples stand out by being considered predominantly acceptable: a supermarket offering a discount to customers holding a loyalty card, a student discount and a quantity discount on bottles of soda. Three more examples are met with neutrality.

Respondents give similar answers regarding two examples for pricing DVDs: price discriminating between a rich and a poor country, and adjusting the price over time. A large majority finds it unacceptable if umbrellas are more expensive when it rains – whether the seller has many umbrellas left is irrelevant to almost anyone.

A majority of the population find some forms of dynamic pricing unacceptable, while those pricing practices are very common and have been around for decades: for instance airlines raising prices when seats are almost sold out, and holiday cottages being more expensive in school holidays. The following section aims to analyse which features make or break the acceptability of price discrimination and dynamic prices.

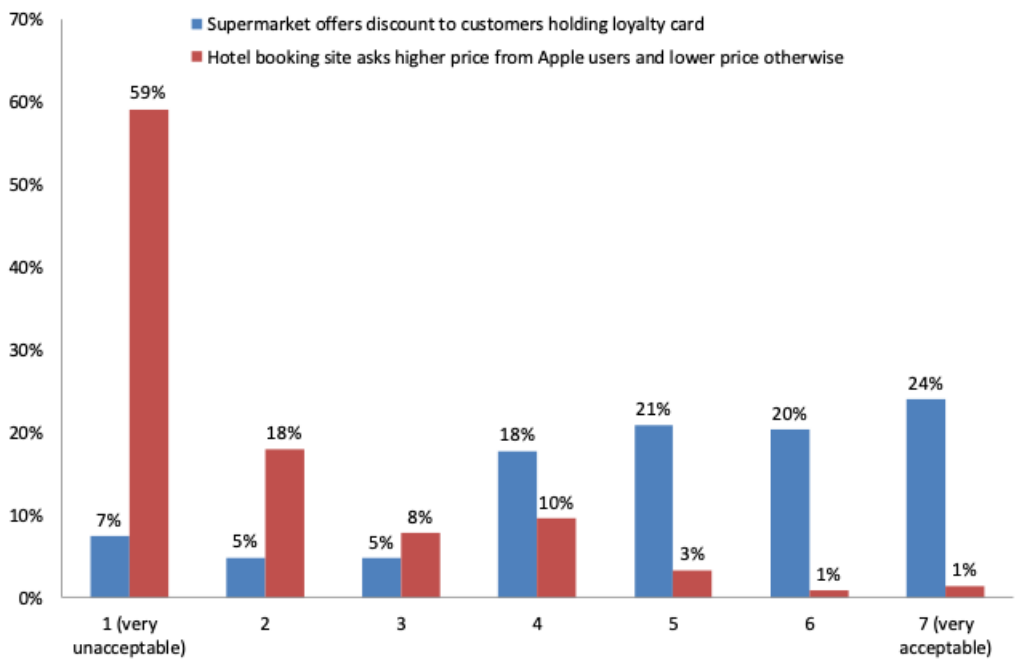

Figure 7: 'Do you find it acceptable when…' (N=1202)

Does everyone have a price? Understanding people's attitude towards online and offline price discrimination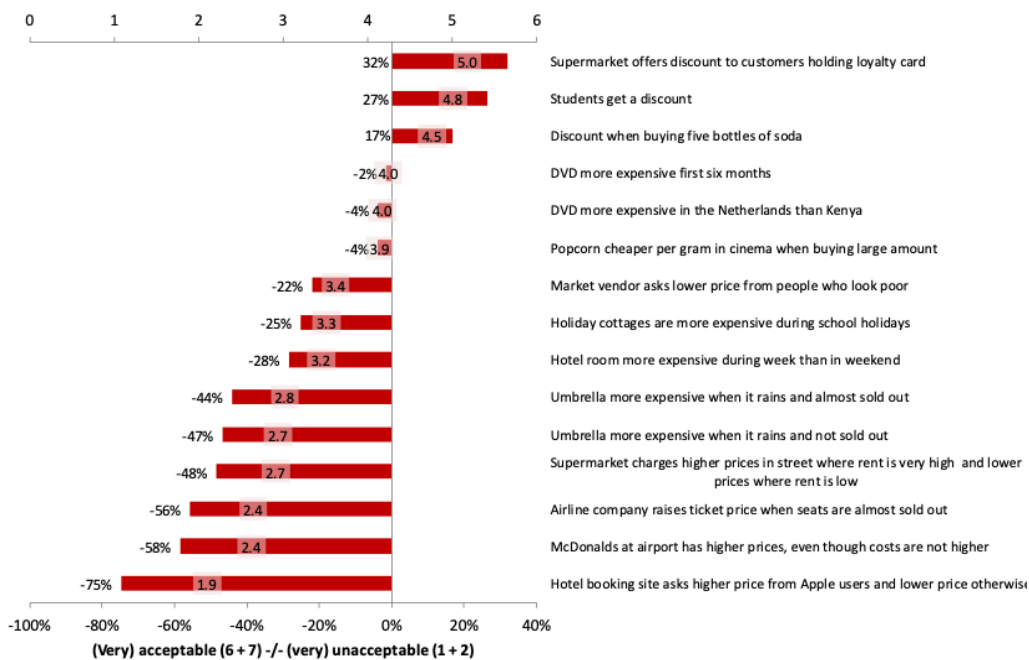

Figure 8: Net acceptability of different forms of price discrimination and dynamic pricing (N=1202)

## 4.4. DISCUSSION OF RESULTS

The first survey shows that more than half of the population claim never to have experienced online price discrimination, while only one in twenty claim to have experienced it often or very often. After being told that online price discrimination is possible, a vast majority finds it unacceptable and unfair. If online price discrimination is presented as a personalised *discount* based on online behaviour, the acceptance is slightly better, but still a majority finds this unacceptable and it makes no difference whether such a discount benefits themselves or others. A majority is worried about paying more than others for a product and about secret price adjustments. In terms of policy measures, an overwhelming majority favours a prohibition and thinks that online stores that engage in online price discrimination should be obliged to inform customers about this.

Why are most people so uncomfortable with online price discrimination and dynamic pricing? Kahneman, Knetsch, & Thaler (1986a, b) unearthed the importance of fairness in transactions, even if fairness is no concern for the neoclassical rational and self-interested *homo economicus*. Indeed, Figure 3 shows that there is hardly any difference between the perceived acceptability and fairness of price discrimination: what is fair is acceptable or *vice-versa*. Moreover, figure 4 showed that respondents' acceptance hardly depends on whether price discrimination favours themselves or others. But what makes certain pricing practices more or less acceptable to people?

To assess this question, Table 1 describes several characteristics of the examples from Figure 8. The second column in Table 1 repeats the net acceptability. The third column indicates whether the example concerns second or third degree price discrimination or neither of these. Five cases are examples of third degree price discrimination, two of second degree price discrimination, and the other examples concern other forms of dynamic pricing.

For these eight other examples, the next column assesses the market dynamics that drive price differentials. In the example of the supermarket charging higher prices in a street where the rent is very high, the underlying idea is that price differences stem from differences in the costs of service. 6 In the other examples, price differences are driven by a demand shift. For these an additional distinction is made between cases in which demand is high and supply is presumed scarce – e.g., holiday cottages during school holidays – and cases in which demand is high and there is sufficient supply – e.g., DVDs in the first six months after release. The last column in Table 1 indicates whether the example is presented in terms of a discount or lower price, or in terms of a higher price, or both.

Table 1: Characteristics of examples of price discrimination and dynamic pricing

| Net acceptability | 2$^{nd}$ or 3$^{rd}$ degree? | Cost or demand shift? | Lower or higher price? |
|---|---|---|---|
| Hotel booking site asks higher price from Apple users and lower price otherwise | -75% | 3$^{rd}$ | Higher and lower |
| McDonalds at airport has higher prices, even though costs are not higher | -58% | Demand, sufficient supply | Higher |
| Airline company raises ticket price when seats are almost sold out | -56% | Demand, scarce supply | Higher |
| Supermarket charges higher prices in street where rent is very high and lower prices where rent is low | -48% | Cost | Higher and lower |
| Umbrella more expensive when it rains and not sold out | -47% | Demand, sufficient supply | Higher |
| Umbrella more expensive when it rains and almost sold out | -44% | Demand, scarce supply | Higher |
| Hotel room more expensive during week than in weekend | -28% | Demand, scarce supply | Higher |
| Holiday cottages are more expensive during school holidays | -25% | Demand, scarce supply | Higher |
| Market vendor asks lower price from people who look poor | -22% | 3$^{rd}$ | Lower |
| Popcorn cheaper per gram in cinema when buying large amount | -4% | 2$^{nd}$ | Lower |
| DVD more expensive in the Netherlands than Kenya | -4% | 3$^{rd}$ | Higher |
| DVD more expensive first six months | -2% | Demand, sufficient supply | Higher |
| Discount when buying five bottles of soda | 17% | 2$^{nd}$ | Lower |
| Students get a discount | 27% | 3$^{rd}$ | Lower |
| Supermarket offers discount to customers holding loyalty card | 32% | 3$^{rd}$ | Lower |

From Table 1 a few observations can be made. Below, we discuss several factors that could help to understand why people approve or disapprove of certain pricing practices. One should bear in mind, however, that the number of cases in Table 1 is too small for regression models linking the characteristics of price discrimination and dynamic pricing to its perceived acceptability. 7

**Second or third degree price discrimination**
First, respondents consider the examples of *second degree* price discrimination (volume discounts) relatively acceptable. Regarding *third degree* price discrimination the picture is more diverse. Some examples are relatively acceptable; others are not. There is one example of *online* third degree price discrimination in this list (a booking site charges more to Apple users). Respondents reject this practice strongly, as they did in more general terms in the first survey. In contrast, respondents generally accept several *offline* examples of third degree price discrimination, even though the examples can be quite intransparent and privacy-invasive, such as a loyalty card in the supermarket, and might be considered unacceptable online. (In the Netherlands, where the survey was conducted, loyalty cards are quite popular; perhaps people see loyalty cards as nothing special because of the cards' popularity.) To conclude, people do not find third degree price discrimination unacceptable *per se*.

**Sufficient or scarce supply**
Second, for cases of *dynamic pricing*, there is *no clear relation* between acceptability and the distinction whether *supply* is *sufficient or scarce*. Instances of sufficient supply score as the most acceptable (DVD in first six months) and the least acceptable (McDonalds at airport) of the cases of dynamic pricing. And in the case of selling umbrellas when it rains, people do not care if supply is scarce or not.

An economist would have more sympathy for higher prices when supply is scarce than when there is ample supply: when supply is scarce, higher prices increase welfare by improving allocative efficiency; when there is ample supply, higher prices primarily extract monopoly rents at the expense of customers. Take the example of holiday cottages: a higher price during school holidays gives people without school-going kids an incentive to go on holidays earlier or later, thereby clearing space for parents that cannot avoid the holiday period. Moreover, higher prices for a scarce good ensure that the good ends up in the hands of the person who values it most in monetary terms.

However, this economic point of view is at odds with a perception of fairness in terms of equal chances for people, regardless of their wealth. Indeed, Kahneman, Knetsch, & Thaler (1986a) demonstrated that most people consider a queue the fairest way to allocate sports tickets, followed by a lottery. Only 4% thought that an auction – the price mechanism – was fairest.

**Framing price discrimination as a discount**
Third, *framing* or phrasing seems to matter significantly: people consider, on average, examples more acceptable if they are presented as a *discount* under some circumstances (or if the examples use the words 'cheaper' or 'lower price'). People find, on average, examples less acceptable if those speak of 'higher prices' or 'more expensive'. Two examples mention both higher and lower prices. Respondents consider these examples highly unacceptable. This suggests that the public only accepts online price discrimination or dynamic pricing if such practices are framed as providing a discount or lower prices.

An economist would reply that discounts and premiums are two sides of the same coin and that you cannot have one without the other – and yet the framing matters to people. On the other hand, the first survey showed that if online price discrimination is presented as a personalised *discount* based on online behaviour, the acceptance is slightly better, but still a majority finds the price discrimination not acceptable. 8

**Loss or regret aversion**

Relatedly, a dislike for price discrimination could also be associated with the concept of *loss* or *regret aversion* (e.g., Loomes & Sugden 1982). People tend to avoid situations that could lead to a loss or regret. Hence, people probably object to a situation in which they would have been offered a better price if they had used a different browser or computer, or deleted their cookies. This ties in with the findings in Figure 6, that a majority of the people are worried that they may pay more than others. It may help explain why people are more accepting towards discounts than towards the other examples: missing out on a discount may feel less like a loss than paying a premium. Yet, the fact that it hardly made a difference whether such discounts benefit respondents themselves or others (Figure 4) suggests that mere personal regret aversion may not suffice to explain people's dislike of price discrimination.

**Transparency**

The examples in Table 1 also differ in aspects that are less objective. One is the *transparency* of the pricing strategy. Student discounts and quantity discounts in the supermarket, for instance, are generally very transparent, while the pricing of airline tickets will be opaque to most people. A likely hypothesis would be that people prefer transparent pricing strategies over ones that are not transparent. Figure 6 provides some evidence in support of this hypothesis by showing that a majority of the people is worried that price adjustments can be made unnoticed. Indeed, very transparent practices such as student discounts and quantity discounts are amongst the most accepted in Table 1. In contrast, it would be hard to observe for people that airline tickets are priced dynamically, or that Apple users pay a premium when booking a hotel online. On the other hand, discounts in the supermarket for loyalty card holders, which is the most accepted strategy in Table 1, can be highly personalised and opaque: customers can rarely observe how much discount other loyalty card holders receive. And it is quite transparent that holiday cottages are more expensive during school vacations and that food prices at most airports are a rip-off. Therefore, transparency does not appear to be the silver bullet to explain the survey outcomes either.

**Captivity**

A last subjective dimension that could drive consumers' attitudes towards different pricing strategies is a feeling of being *captive*; of having no real choice. In general, anyone could get a loyalty card from a supermarket to become eligible for discounts and one can decide for oneself to go for a quantity discount and buy several bottles of soda or a large bucket of popcorn. And if one wants to pay less for a DVD, it is normally wise to wait a couple of months.

In contrast, behind the security of an airport, people cannot walk a few blocks to a cheaper competitor. And parents with young kids typically have to book their vacation during school holidays. Such captivity may be a key to some of the differences in acceptability. But there are anomalies for this explanation as well, such as the supermarket charging higher prices in a street where the rents are high, even though people could walk a few blocks to find a cheaper store.

## 5. DEMOGRAPHIC PATTERNS

Lastly, we analyse how demographic factors drive the perceived acceptability of different examples of price discrimination. To this end, we estimated a simple regression model (OLS) for the sum-score of the reported acceptability of all fifteen questions. This sum-score ranges from 15 (i.e., for someone scoring '1' for totally unacceptable on each of the fifteen questions) to 105 (fifteen times '7' for totally acceptable). Table 2 summarises the model outcomes. All coefficients are statistically significant at a 95% confidence level. 9

Table 2: Demographic factors driving acceptance of pricing strategies

| Coefficient | | St. Error | Sig. |
|---|---|---|---|
| Constant | 62.40 | 2.82 | 0.00 |
| Gender | -5.10 | 0.91 | 0.00 |
| Age | -0.27 | 0.03 | 0.00 |
| Monthly net household income (× €1000) | 0.66 | 0.31 | 0.03 |
| Education level | 2.61 | 0.32 | 0.00 |

A higher score on the dependent variable stands for 'more acceptable'. Therefore, the results read as follows:

- The negative sign for gender (1 = male, 2 = female) implies that men find price discrimination and dynamic pricing strategies more acceptable than women.
- The negative sign for age leads to the conclusion that younger people have a higher acceptance of these cases than older people.
- The positive sign for household income implies a positive correlation between income and the acceptance of price discrimination and dynamic pricing strategies.
- The positive coefficient for education level implies that the higher a person is educated, the more accepting he or she is towards these cases.

In sum, young, highly educated males in higher income groups have the highest acceptance of price discrimination and dynamic pricing strategies, while older, lower educated females in lower income groups have the lowest acceptance.

One can hypothesise that the link with education is driven by a better understanding of the business interests of pricing practices or even the positive effect such strategies may have on allocation and welfare. Even a mere understanding of what is going on may cause greater acceptance. This hypothesis is related to the point of transparency and even captivity made in the previous section: pricing may be more transparent to people who have a better understanding of what is going on. The link with age on all four models could have to do with in general a better understanding of digital developments, but on the other hand, many of the examples given are offline.

The positive correlation between acceptance and income is perhaps the most interesting here, as higher income groups may expect to be the 'victim' of price discrimination, in the sense that they can reasonably expect to be the ones to pay the higher prices. It would be tempting to conclude that perceived fairness trumps self-interest here. However, a more down-to-earth explanation could be that richer people simply care less, or in fact benefit from the beneficial allocative effects of price discrimination and dynamic pricing.

# 6. CONCLUDING THOUGHTS

In this article we analysed two surveys on online price discrimination and various forms of dynamic pricing. Such pricing practices fit in a broader trend towards data-driven or algorithmic personalisation of services. We tried to gain a better understanding of the attitudes of the general public towards it and the underlying factors.

In general, we find that more than half of the population claim never to have experienced online price discrimination, while only one in twenty claims to have experienced it often or very often. Nevertheless, an overwhelming majority considers online price discrimination unacceptable and unfair. If such price discrimination is presented as a personalised *discount* based on online behaviour, the acceptance is slightly better, but still a majority finds this unacceptable and it makes no difference whether such a discount benefits respondents or others. A majority is worried about paying more than others for a product and about secret price adjustments.

Based on fifteen specific examples of price discrimination and dynamic pricing, we discuss several factors that could help to understand why people approve or disapprove of certain pricing practices. In general, acceptance increases with income and education level, and decreases with age. Men are on average more accepting to price discrimination and dynamic pricing than women. People dislike some pricing strategies that have been commonly applied for decades, such as higher prices for holiday cottages during school holidays and higher prices for airplane tickets when a flight is almost full. In addition, the following observations can be made:

- Examples of second degree price discrimination (quantity discounts) are comparably well-accepted. For third degree price discrimination (between groups of customers), the pattern is mixed. People do not find third degree price discrimination unacceptable *per se*.
- Whether supply is sufficient or scarce hardly matters to people for their acceptance of dynamic pricing. From a welfare-economic perspective, dynamic pricing is much more defendable when supply is scarce.
- People are much more willing to accept price discrimination and dynamic pricing if it is framed as a discount.
- Loss or regret aversion, a lack of transparency, and a feeling of being captive may be other factors that drive people's negative attitude towards price discrimination and dynamic pricing.

All of these factors help explain some of the patterns seen in the attitudes towards price discrimination, but in each case, there are anomalies too.

A vast majority of the population would favour a prohibition of online price discrimination and thinks that online stores that engage in online price discrimination should be obliged to inform customers about this. However, a ban on online price discrimination may not be wise from a welfare-economic perspective. Besides, people also exhibit a great dislike of several offline forms of price discrimination and dynamic pricing.

From a legal perspective, it also makes sense to explore lighter measures. Transparency requirements could be worth considering, in particular since transparency seems to be a factor that affects people's acceptance, while online it is much easier to price discriminate secretly. We argued in earlier work (Zuiderveen Borgesius & Poort, 2017) that the GDPR applies when a web store personalised prices. If a company uses personal data to recognise customers and to adapt prices, the company must disclose this. However, the lack of web stores that state clearly that they engage in such price discrimination indicates that transparency is still far away.

The authors would like to thank Federico Morando, Joe Karaganis, and the journal editors for their valuable suggestions. We would also like to thank Claes de Vreese, Natali Helberger, Sophie Boerman, and Sanne Kruikemeijer.

Does everyone have a price? Understanding people's attitude towards online and offline price discrimination

**REFERENCES**

Amazon. (2000, September 27). *Amazon.com Issues statement regarding random price testing* [Press release]. Retrieved from
http://phx.corporate-ir.net/phoenix.zhtml?c=176060&p=irol-newsArticle_Print&ID=502821

Armstrong, M. (2006). Recent developments in the economics of price discrimination. In R. Blundell, W. Newey, & T. Persson (Eds.), *Advances in economics and Econometrics, Theory and Applications, Ninth World Congress* (Vol. II, pp. 97-141). Cambridge; New York: Cambridge University Press. Preprint available at http://discovery.ucl.ac.uk/14558/

BBC News. (2000, September 8). *Amazon's old customers 'pay more'.* Retrieved from:
http://news.bbc.co.uk/2/hi/business/914691.stm

ePrivacy Directive (2009). Directive 2002/58/EC on Privacy and Electronic Communications, last amended in 2009.

Executive Office of the President of the United States (2015). *Big data and differential pricing.* Retrieved from:
https://obamawhitehouse.archives.gov/sites/default/files/whitehouse_files/docs/Big_Data_Report_Nonembargo_v2.pdf

General Data Protection Regulation (2016). Regulation on the protection of natural persons with regard to the processing of personal data and on the free movement of such data, and repealing Directive 95/46/EC (EU 2016/679).

Geo-Blocking Regulation 2018/302 (2018) Regulation (EU) 2018/302 of the European Parliament and of the Council of 28 February 2018 on addressing unjustified geo-blocking and other forms of discrimination based on customers' nationality, place of residence or place of establishment within the internal market and amending Regulations (EC) No 2006/2004 and (EU) 2017/2394 and Directive 2009/22/EC, OJ L 60I 2 March 2018, p. 1-15

Hannak, A., Soeller, G., Lazer, D., Mislove, A., & Wilson, C. (2014). Measuring price discrimination and steering on e-commerce web sites. In *Proceedings of the 2014 Conference on Internet Measurement Conference* (pp. 305-318). New York: ACM. doi: 10.1145/2663716.2663744

Kahneman, D., Knetsch, J. L. and Thaler, R. H. (1986a). Fairness and the assumptions of economics. *Journal of Business, 59*(4), S285-S300. Retrieved from
https://www.jstor.org/stable/2352761

Kahneman, D., Knetsch, J. L. & Thaler, R. H. (1986b). Fairness as a constraint on profit seeking: Entitlements in the market. *The American Economic Review, 76*(4), 728-741. Retrieved from
https://www.jstor.org/stable/1806070

Loomes, G., & Sugden, R. (1982). Regret theory: An alternative theory of rational choice under uncertainty. *The Economic Journal, 92*(368), 805-824. doi:10.2307/2232669

Mikians, J., Gyarmati, L., Erramilli, V., & Laoutaris, N. (2013). Crowd-assisted Search for Price Discrimination in e-Commerce: First Results. In *Proceedings of the Ninth ACM Conference on Emerging Networking Experiments and Technologies* (pp. 1–6). New York: ACM. doi:10.1145/2535372.2535415 Preprint available at https://arxiv.org/abs/1307.4531v1

Neppelenbroek, E. D. C. (2016). Wat de gek ervoor geeft: Big data en de bescherming van de contractuele wederpartij bij prijsdiscriminatie ['big data and protection of the contract party in the context of price discrimination']. *Weekblad Voor Privaatrecht, Notariaat En Registratie, 147*(7110), 443-452.

Odlyzko, A. (2009). Network neutrality, search neutrality, and the never-ending conflict between efficiency and fairness in markets. *Review of Network Economics 8*(1), 40-60. doi:10.2202/1446-9022.1169

Pigou, A.C (1932). *The economics of welfare*. London: Macmillan & Co.

Schulte-Nölke, H., Zoll, F., Macierzyńska-Franaszczyk, E., Stefan, S., Charlton, S., Barmscheid, M., & Kubela, M. (2013). *Discrimination of Consumers in the Digital Single Market* (Study No. IP/A/IMCO/ST/2013-03, PE 507.456). Brussels: European Parliament, Directorate-General for Internal Policies, Policy Department A: Economic and Scientific Policy. Retrieved from http://www.europarl.europa.eu/meetdocs/2014_2019/documents/imco/dv/discrim_consumers_/discrim_consumers_en.pdf

Steppe, R. (2017). Online price discrimination and personal data: A General Data Protection Regulation perspective. *Computer Law & Security Review*, *33(6)*, 768-785. doi:10.1016/j.clsr.2017.05.008

Stigler, G.J. (1987). *Theory of price*. (Fourth Ed.). New York, USA: Macmillan

Turow, J., Feldman, L., & Meltzer, K. (2005). Open to exploitation: America's shoppers online and offline (Working Paper). Philadelphia: Annenberg Public Policy Center of the University of Pennsylvania. Retrieved from: http://repository.upenn.edu/asc_papers/35

Turow J., King, J., Hoofnagle, C.J., Bleakley, A. & Hennessy, M. (2009). *Americans Reject Tailored Advertising and Three Activities That Enable It* (Technical Report). Philadelphia; Berkeley: Annenberg School for Communication; Berkeley School of Law. Available at https://repository.upenn.edu/asc_papers/137/

Valentino-Devries, J., Singer-Vine, J., & Soltani, A. (2012, December 23). Websites vary prices, deals based on users' information. *Wall Street Journal*. Retrieved from: http://online.wsj.com/article/SB10001424127887323777204578189391813881534.html

Varian H.R. (1989). Price discrimination. In R. Schmalensee, & R.D. Willig (Eds.), *Handbook of Industrial Organization* (Vol. 1, pp. 597-654). Amsterdam; New York: Elsevier. doi:10.1016/S1573-448X(89)01013-7

Vissers, T., Nikiforakis, N., Bielova, N., & Joosen, W. (2014). Crying wolf? On the price discrimination of online airline tickets. Presented at the 7th Workshop on Hot Topics in Privacy Enhancing Technologies (HotPETs 2014). Retrieved from https://hal.inria.fr/hal-01081034/document

Zuiderveen Borgesius, F. (2015). *Online Price Discrimination and Data Protection Law* (Research Paper No. 2015–32). Amsterdam: University of Amsterdam Law School. Retrieved from http://hdl.handle.net/11245/1.506283

Zuiderveen Borgesius, F. & Poort, J. (2017). Online Price Discrimination and EU Data Privacy Law. *Journal of Consumer Policy 40*(3), 347-366. doi:10.1007/s10603-017-9354-z

**FOOTNOTES**

1. This section is based on Zuiderveen Borgesius and Poort (2017).

2. An economically more proper definition of price discrimination by Stigler is "the sale of two or more similar goods at prices that are in different ratios to marginal costs." (Stigler, 1987: p. 210). Under this definition, price differences that purely stem from cost differences would not qualify as price discrimination. Versioning might qualify as a result of the rather vague word 'similar'.

3. Sometimes, individual prices that do not equal each individual's willingness to pay are also referred to as first degree price discrimination. This use of the term is confusing and avoided in this paper.

4. The LISS panel (Longitudinal Internet Studies for the Social sciences) is a panel for academic research purposes, administered by CentERdata (Tilburg University, The Netherlands). The panel consists of 4,500 households, comprising 7,000 individuals. It is based on a true probability sample of households drawn from the population register by Statistics Netherlands. Households that could not otherwise participate are provided with a computer and internet connection.

5. All surveys were in Dutch; questions have been translated into English for this paper.

6. It cannot be ruled out that some respondents misinterpreted this example as suggesting rents for houses are high, which would translate to a larger average purchasing power (a demand shift).

7. Moreover, the responses to many of the cases correlate significantly at an individual level (mostly significant at a 0.01 level). This suggests that many of these questions might measure closely related concepts or attitudes. In such situations, exploratory *factor analysis* can be useful to extract the key constructs that underlie the cases in Figure 8 and Table 1 and to shed light on patterns or similarities that do not meet the eye. Using Principal Axis Factoring, a three factor solution was found to explain 60% of the variance in the data set. The rotated factor scores per case follow the net acceptability in Figure 8 closely. This implies that factor analysis yields no additional information about what drives people's attitude towards price discrimination or dynamic pricing and for this reason, the results are not reported in this paper. They can be obtained from the authors.

8. These outcomes are largely in line with the study by Turow at al. (2009), in which 78% indicated they did not want discounts tailored based on what they did on other websites.

9. Separate models have been estimated on the three factors referred to in footnote 7 and yield highly similar outcomes on these demographic variables.